\documentclass[aps,showpacs,twocolumn,superscriptaddress]{revtex4-1}

\usepackage{graphicx}
\usepackage{dcolumn}
\usepackage{bm}
\usepackage{amsmath}

\begin{document}

\preprint{}

\title{Detecting and interpreting distortions in hierarchical organization of complex time series}

\author{Stanis\l{}aw~Dro\.zd\.z}
\affiliation{Institute of Nuclear Physics, Polish Academy of Sciences,
Krak\'ow, Poland.}
\affiliation{Faculty of Physics, Mathematics and Computer Science, Cracow
University of Technology, Krak\'ow, Poland.}

\author{Pawe\l{}~O\'swi\c ecimka}
\email{pawel.oswiecimka@ifj.edu.pl}
\affiliation{Institute of Nuclear Physics, Polish Academy of Sciences, Krak\'ow, Poland.}

\date{\today}

\begin{abstract}

Hierarchical organization is a cornerstone of complexity and multifractality constitutes its
central quantifying concept. For model uniform cascades the corresponding singularity spectra
are symmetric while those extracted from empirical data are often asymmetric.
Using the selected time series representing such diverse phenomena like price changes
and inter-transaction times in the financial markets, sentence length variability
in the narrative texts, Missouri River discharge and Sunspot Number variability as examples,
we show that the resulting singularity spectra appear strongly asymmetric, more often left-sided
but in some cases also right-sided. We present a unified view on the origin of such effects and indicate
that they may be crucially informative for identifying composition of the time series.
One particularly intriguing case of this later kind of asymmetry is detected in the daily reported
Sunspot Number variability. This signals that either the commonly used famous Wolf formula distorts
the real dynamics in expressing the largest Sunspot Numbers or, if not, that their dynamics is governed
by a somewhat different mechanism.

\end{abstract}


\maketitle

Multi-scale approach~\cite{halsey86,mandelbrot89,muzy94} aims at bridging the wide range of time and length
scales that are inherent in a number of essential processes in complex natural phenomena.
Multifractality constitutes a leading concept towards quantifying the related characteristics~\cite{kwapien12}.
By now it finds applications in essentially all areas of the scientific activity including
physics~\cite{muzy08,subramaniam08},
biology~\cite{ivanov99,makowiec09,rosas02},
chemistry~\cite{stanley88,udovichenko02}, geophysics~\cite{witt13,telesca05},
economics~\cite{ausloos02,calvet02,turiel05,drozdz10,oswiecimka08,zhou09,bogachev09,su09},
hydrology~\cite{koscielny06}, atmospheric physics~\cite{kantelhardt06},
quantitative linguistics~\cite{ausloos12,grabska12},
behavioral sciences~\cite{ihlen13},
music~\cite{jafari07,oswiecimka11},
and even ecological sciences~\cite{stephen13}.

At present the most efficient, numerically stable and precise~\cite{oswiecimka06} method to quantify 
multifractality is based on the Multifractal Detrended Fluctuation Analysis (MFDFA)~\cite{kantelhardt02}.
Accordingly, for a discrete signal ${x(i)}_{i=1,...,N}$ one starts with the signal profile
$X(j) = \sum_{i=1}^j{(x(i)-<x>)}, \ j = 1,...,N$, where $<...>$ denotes averaging over
all $i$'s. Then one divides the $X(j)$ into $M_s$ non-overlapping segments of
length $s$ ($s < N$) starting from both the beginning and the end of the
signal (there are thus in total $2 M_s$ segments). For each segment a local trend is estimated
by fitting an $l$th order polynomial $P_{\nu}^{(l)}$, which is then subtracted
from the signal profile. For the so-detrended signal a local variance
$F^2(\nu,s)$ in each segment $\nu$ is calculated for the scale variable $s$.
Finally, by averaging $F^2(\nu,s)$ over all segments $\nu$ one calculates the $q$th order
fluctuation function:
\begin{equation}
F_q(s) = \bigg\{ \frac{1}{2 M_s} \sum_{\nu=1}^{2 M_s} [F^2(\nu,s)]^{q/2} \bigg\}^{1/q},
\label{ffunction}
\end{equation}
and $q \in \mathbf{R}$. The optimal range is $q \in [-4,4]$~\cite{drozdz09}.

The scaling behavior of $F_q(s) \sim s^{h(q)}$ indicates fractal structure with the singularity spectrum $f(\alpha)$~\cite{halsey86}
\begin{equation}
f (\alpha) =q [\alpha- h (q)] + 1
\label{falpha}
\end{equation}
for $\alpha=h(q)+q h'(q)$. If $h(q)= {\rm const}$, the signal is monofractal. A nontrivial $q$-dependence
of $h(q)$ indicates its more convoluted fractal organization termed multifractal.
For model multifractal series $f(\alpha)$ typically assumes shape of an inverted symmetric parabola
while the empirical ones often develop asymmetries.
Here we explore this issue and document that asymmetries in $f(\alpha)$ may provide important information
about the series organization.

\begin{figure*}
\includegraphics[width=0.74 \textwidth, height=0.57 \textwidth]{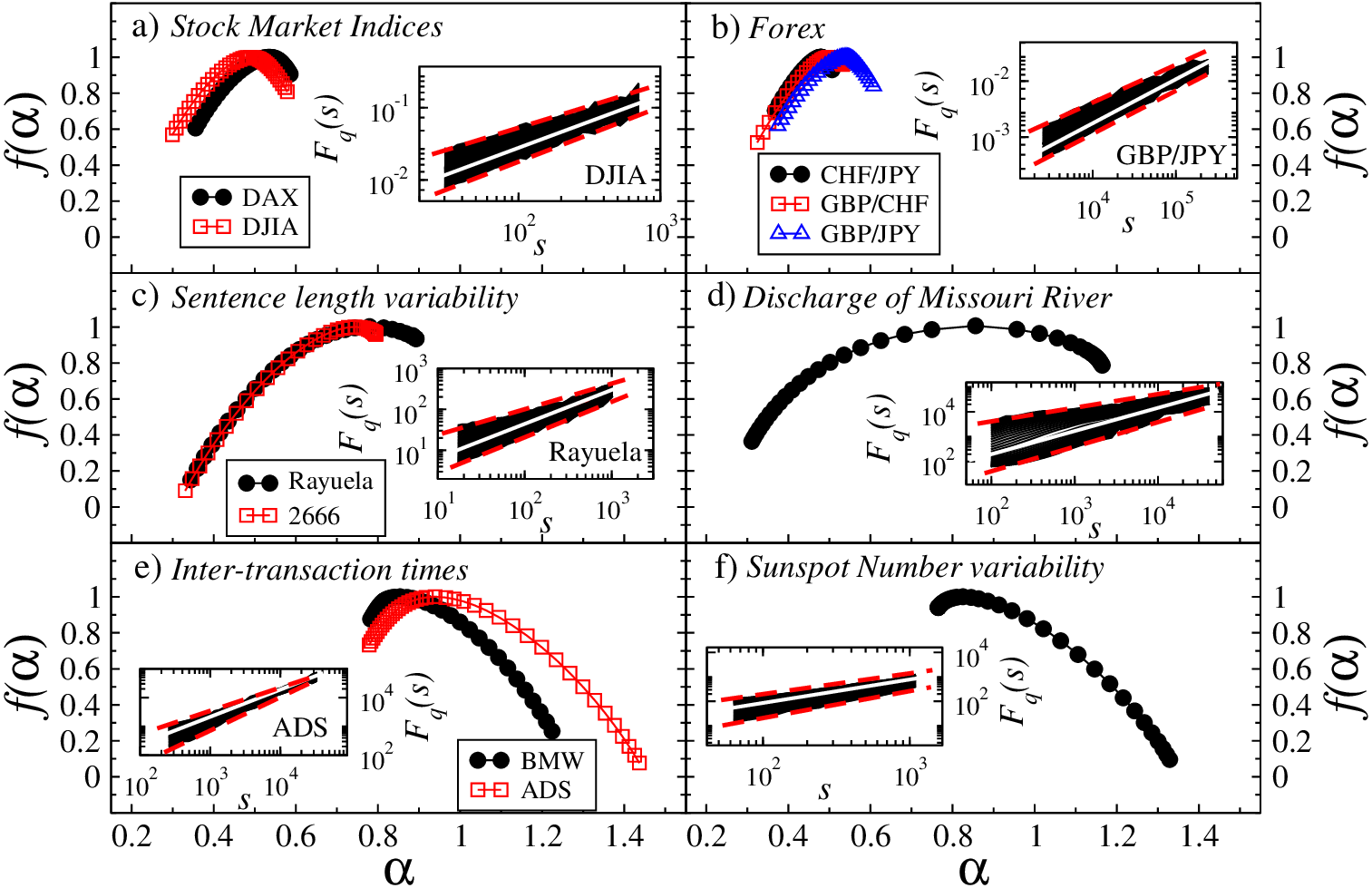}
\caption{(Color online) Singularity spectra $f(\alpha)$ for (a) daily returns of the two (DJIA and DAX) stock market indices
over the period January 12, 1990 - October 12, 2013 (5881 data points),
(b) minutely Forex returns on selected pairs (CHF/JPY,  GBP/CHF, GBP/JPY) of currencies over the period
9:00 pm January 2, 2004 - 9:00 pm 20 March, 2008 (1703520 data points each), (c) the series representing length,
expressed in terms of the number of words, of the consecutive sentences in the two narrative texts:
{\it Rayuela} by Julio Cort\'{a}zar (9848 sentences) and {\it 2666} by Roberto Bola\~{n}o (21319 sentences), (d) hourly
changes of the Missouri River discharge
(in cubic feet per second) at Waverly, taken from the U.S. Geological Survey database~\cite{water07} over the period
October 1, 1987 - October 1, 2007 (213344 data points), (e) inter-transaction intervals over the period
November 28, 1997 - December 31, 1999, for the two German stock market companies:
ADS (167208 data points) and BMW  (173664 data points) (f) daily Sunspot Number over the period January 1, 1900 -
June 30, 2014 (41817 data points); Source: WDC-SILSO, Royal Observatory of Belgium, Brussels~\cite{sunspot14}.
The insets show the corresponding $F_q(s)$ for $q \in [-4,4]$ with the white strip indicating
$q=0$. The detrending polynomial used, as optimal, is of the second order. In case (f) the upper limit of the scaling
parameter $s$ is less than half of the 11 years solar activity cycle which makes this procedure unbiased by the periodic
trend.}
\label{fig1}
\end{figure*}

Spectacular examples, largely novel, of unquestioned asymmetric multifractal spectra, generated using
MFDFA algorithm, for signals representing diverse, mutually remote areas are shown in Fig.~\ref{fig1}.
They are seen to even display two distinct kinds of asymmetry.
The DAX and DJIA stock market indices, the Forex market represented by the three exchange rate
pairs (CHF/JPY, GBP/CHF and GBP/JPY), the series representing the sentence length variability in some
selected literary texts (here {\it Rayuela} by Julio Cort\'{a}zar and {\it 2666} by Roberto Bola\'{n}o;
most of the world famous literary works are essentially monofractal in the sentence length variability
and just a dozen or so are convincingly multifractal and the two considered here at the same time belong
to those few that develop the most asymmetric $f(\alpha)$)
and the discharge of Missouri River~\cite{water07} show pronounced left-sided asymmetry.
A rarer, right-sided asymmetry is displayed by the other two examples,
the stock market company (here ADS and BMW from DAX) inter-transaction times
and the Sunspot Number variability~\cite{bray79}. 
In order to estimate a possible additional contribution to asymmetry of the potential 'oscillating
singularities'~\cite{seuret06}  we also applied the so-called wavelet leaders algorithm~\cite{lashermes08,serrano09}
to all these data but detected no significant signals of such oscillations. 

The left side of $f(\alpha)$ is determined by the positive $q$-values, which filter out larger events
and the opposite applies to its right side.
Hence, asymmetry in $f(\alpha)$ signals non-uniformity of the underlying cascade. The (a)-(d) cases in
Fig.~\ref{fig1} are thus seen to be more multifractal in arrangement of the large events and far less such in
the small ones. In some of those cases (CHF/JPY, GBP/CHF or 2666) the right-side of $f(\alpha)$
contracts so strongly that it indicates essentially a monofractal character of the corresponding small fluctuations.
This suggests that the entire signal can be considered a mixture of the large scale homogeneous multiplicative
cascade, giving thus rise to a pronounced left wing in $f(\alpha)$, and of the small scale noise-like background,
shrinking the corresponding right wing in $f(\alpha)$. Such a noise-like component in empirical data is not very
unusual as it may for instance originate from the measurement uncertainty or from some coarse graining that
effectively affects more just the small scale fluctuations.
The reverse kind of asymmetry - right-sided - applies, however, to the cases (e) and (f), and here the situation
is more intriguing and unusual since it indicates that pronounced multifractality operates on small scale fluctuations
while the dynamics of large scale fluctuations is much poorer in this respect. That the stock market company
inter-transaction intervals may be governed by such a dynamics is conceivable. Small intervals occur between
transactions grouped within the same clusters of an enhanced volatility on this particular company and may thus
be strongly, also nonlinearly correlated. The distances between such in time more separated clusters
of activity, thus larger inter-transaction intervals connecting them, may be less
correlated~\cite{drozdz09}.

More difficult to interpret - thus even more interesting - is the analogous result for the Sunspot Number
variability. The smaller fluctuations (seen through $q < 0$) develop a broad right-sided $f(\alpha)$
which suggests their cascade-like hierarchical organization. At the same time the large ones, filtered out
by $q > 0$, show at most a remnant of multifractality. They thus do not belong to the same hierarchy or,
at best, strongly distort it.
This observation may either demand revision of the famous Wolf formula ($R=k*(N_s + 10 N_g)$, $N_s$ - number of spots,
$N_g$ - number of groups)~\cite{wolf1859} commonly used for expressing the Sun activity in terms of the number of spots
or may indicate that the corresponding large and small activities are governed by a somewhat different cascading mechanisms.

The quality of scaling can be assessed from the insets to the corresponding panels of Fig.~\ref{fig1} which show
the $F_q(s)$ dependencies on $s$. In (a) - (d) one in addition sees that the scaling exponents (slope of $F_q(s)$)
visibly depend on $q$ for $q > 0$  while very weakly for $q < 0$. The opposite is true for (e) and (f) cases.
Naturally, this correlates with orientations of asymmetries seen in $f(\alpha)$.

\begin{figure}
\includegraphics[width=0.45\textwidth, height=0.37 \textwidth]{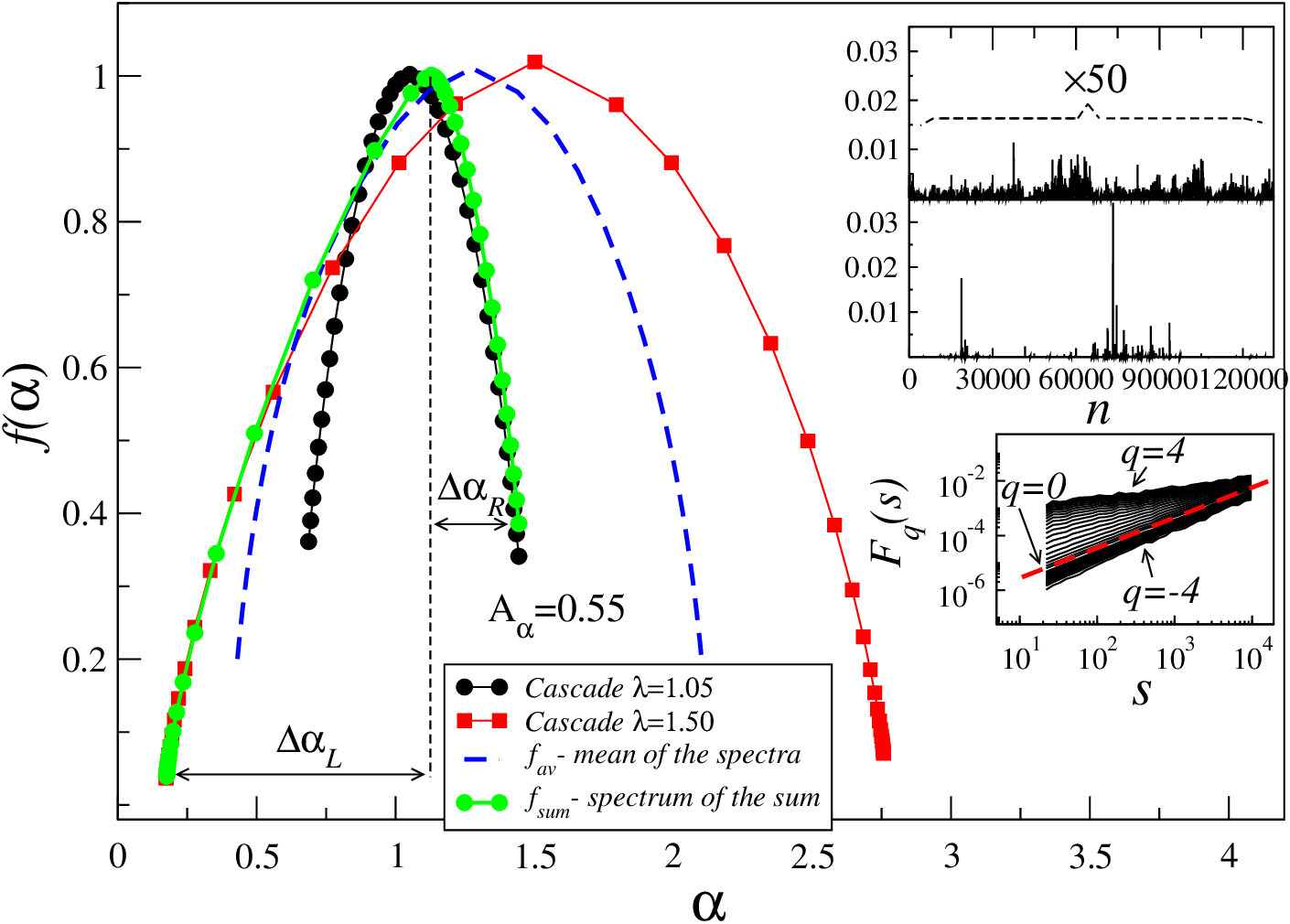}
\caption{(Colore online) Examples of $f(\alpha)$ for the two binomial cascades with $\lambda = 1.05$
and $\lambda = 1.5$, the average $f_{av}(\alpha)$ of such two $f(\alpha)$'s and $f_{sum}(\alpha)$ calculated
for the sum of these two cascades. The upper two insets show explicit spreads of these two cascades.
To better reflect proportions in the same scale the original values of $\lambda = 1.05$ cascade
are multiplied by a factor of 50. The lowest inset displays $F_q(s)$ functions for the sum of both cascades.}
\label{fig2}
\end{figure}

The most likely origin and thus interpretation of the above effects can transparently be illustrated by using
the model cascades. Here we use the binomial cascade with multipliers whose logarithms are drawn from the
Gaussian $N(\lambda,1)$  distribution. Examples of $f(\alpha)$ for two such cascades with $\lambda = 1.05$
and $\lambda = 1.5$, the average $f_{av}(\alpha)$ of such two $f(\alpha)$'s and $f_{sum}(\alpha)$ calculated
for the sum of these two cascades are shown in the main panel of Fig.~\ref{fig2}. While $f_{av}(\alpha)$ is symmetric,
the $f_{sum}(\alpha)$ is much narrower and already strongly asymmetric. To quantify these characteristics
we introduce the asymmetry parameter
\begin{equation}
A_{\alpha} = ({\Delta \alpha}_L - {\Delta \alpha}_R) / ({\Delta \alpha}_L + {\Delta \alpha}_R)
\label{ap}
\end{equation}
where ${\Delta \alpha}_L = {\alpha}_0 - {\alpha}_{min}$ and ${\Delta \alpha}_R = {\alpha}_{max} - {\alpha}_0$
and $\alpha_{min}$, $\alpha_{max}$, $\alpha_0$ denote the beginning and the end of $f(\alpha)$ support, and the
$\alpha$ value at maximum of $f(\alpha)$ (which corresponds to $q=0$), respectively.
A simple superposition of only two such stochastic binomial cascades with
different parameters produces strongly left-sided asymmetric ($A_{\alpha}=0.55$) multifractal spectrum $f(\alpha)$,
resembling the corresponding empirical ones from Fig.~\ref{fig1}. It is instructive to see that the left side of
$f_{sum}(\alpha)$ coincides with $f(\alpha)$ for the $\lambda = 1.5$ cascade while its right side with the one for
$\lambda = 1.05$ cascade. The origin of this result becomes clear from the relative proportions of the size of
fluctuations in these two individual cascades as seen in the upper two panels in Fig.~\ref{fig2}. The $\lambda = 1.05$
fluctuations are on average about two orders of magnitude smaller than the $\lambda = 1.5$ ones.
Thus in their sum the small fluctuations are dominated by $\lambda = 1.05$ and filtered out by $q < 0$ which corresponds
to the right side in $f_{sum}$ and the opposite applies to large fluctuations.
An equivalent way to see this effect is through $F_q(s)$ defined by Eq.~\ref{ffunction}
and shown in the lowest panel of Fig.~\ref{fig2}. Indeed, all $F_q(s)$ scale with $s$
but spread much more for $q > 0$, thus width on the left side of $f_{sum}(\alpha)$ is larger.

\begin{figure}
\includegraphics[width=0.45\textwidth, height=0.37 \textwidth]{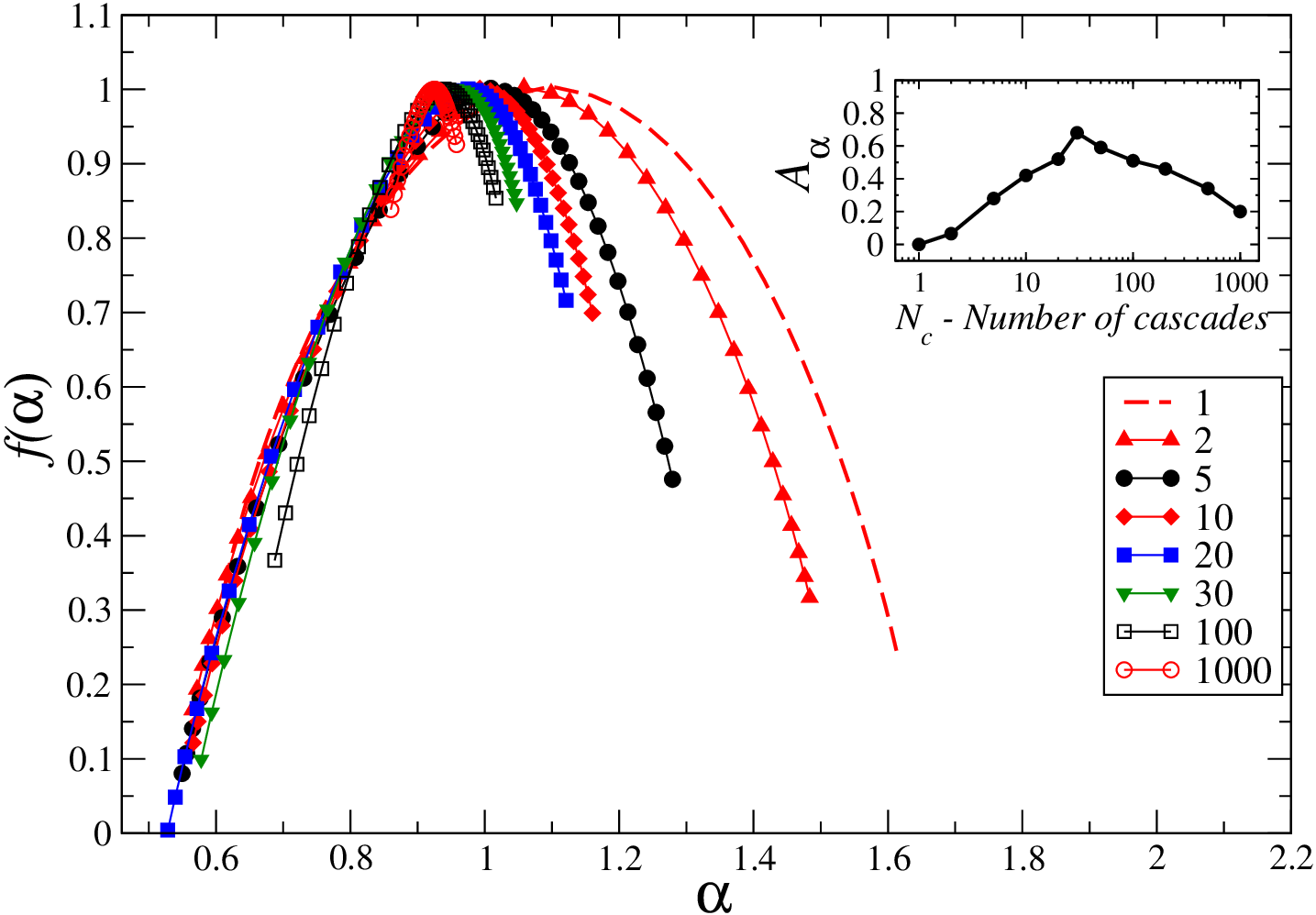}
\caption{(Color online) Singularity spectra of $f(\alpha)$ for an increasing number of the superimposed binomial cascades
generated independently but all with the same $\lambda = 1.1$. Inset illustrates how the corresponding
asymmetry parameter $A_{\alpha}$ evolves.}
\label{fig3}
\end{figure}

Similar effect of left-sided asymmetry one even obtains by superimposing cascades generated with the same $\lambda$.
The result for an increasing number $N$ of independently drawn and then superimposed cascades with $\lambda = 1.1$
is displayed in Fig.~\ref{fig3}. Here however more components are needed to reach the same degree of asymmetry as in the
Fig.~\ref{fig2} case.
The left-sided asymmetry originates here from the fact that small fluctuations are more abundant in such individual
cascades therefore summing up an increasing number of them one sooner approaches the white noise limit, i.e., destroys
the original hierarchical organization on the level of small fluctuations than on the level of the large ones.
The right side of the resulting $f(\alpha)$ contracts sooner to a monofractal with an increasing $N$ than does the
corresponding left side. This can be seen from Fig.~\ref{fig3}, especially from the inset which shows the $A_{\alpha}$
dependence on $N$. After the initial sharp increase it reaches maximum at around $N = 30$ with $A_{\alpha} \approx 0.7$
and then slowly starts decreasing. For $N=1000$ the $f(\alpha)$ already is much closer to a monofractal with $A_{\alpha} \approx 0.2$.

\begin{figure}
\includegraphics[width=0.45\textwidth, height=0.37 \textwidth]{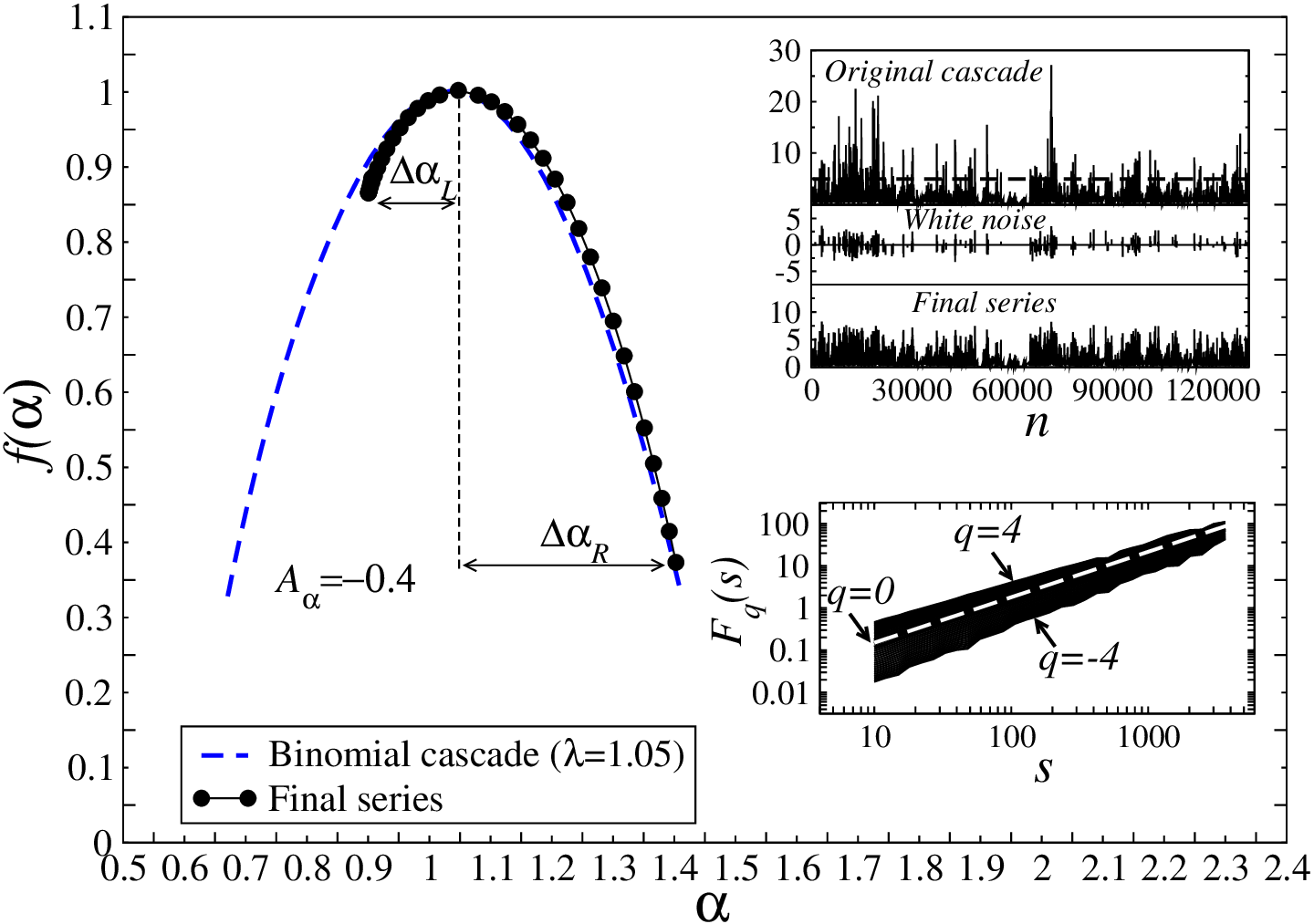}
\caption{(Color online) Singularity spectrum $f(\alpha)$ (i) for $\lambda = 1.05$ binomial cascade and (ii) for the one obtained from
it by setting a threshold of 4.7, then from all events that exceed this threshold value subtracting the over-threshold
value (which affects $1\%$ of all events) and finally by adding consecutively to those affected one number drawn from
a normal distribution, one independently to each. Upper inset, from top to bottom, shows the original binomial cascade
with the truncation threshold (dashed line), the white noise added and the resulting final series. Lower inset displays
the $q$-dependence of the fluctuation functions $F_q(s)$ for the so-obtained final series.}
\label{fig4}
\end{figure}

While the left-sided multifractal asymmetry easily emerges when superimposing cascades, the right-sided
one is more peculiar to model.
From an algorithmic perspective it implies just a reverse,i.e., a more uniform hierarchical organization on the level of smaller fluctuations and more noise-like behavior of the large fluctuations. Such an interpretation indicates construction of a series
obeying condition of this kind.
An example is displayed in Fig.~\ref{fig4}. This is the binomial cascade generated from the same $\lambda = 1.05$ model
as before, but now a top section of the largest events is randomized as follows.
$1\%$ of the largest events is selected which in this case corresponds to those that exceed a threshold of $T = 4.7 \sigma$.
Then, their values are replaced by the sum of $T$ and of a number drawn independently for each event from $N(0,1)$.
The net result of such a construction preserves organization of the original small fluctuations and randomizes the largest
ones. The so obtained series is displayed in the upper inset of Fig.~\ref{fig4} ({\it final series}), together with the two
components that are used to form it. The corresponding $F_q(s)$ functions (lower panel in Fig.~\ref{fig4}) become now more
spread for $q < 0$, thus the $f(\alpha)$ spectrum becomes right-sided ($A_{\alpha} = -0.4$), which confirms the demanded reversed organization. Of course, by playing with $T$ or with noise added one may modulate the degree of this right-sided asymmetry.

In conclusion the effects of asymmetry in singularity spectra $f(\alpha)$ of the time series representing complex systems may not just be numerical artifacts as they seem to be treated in the literature on the subject, but instead may contain genuine information about the composition of such series.They then indicate at least two different regimes in the multifractal scaling.
The left-sided asymmetry corresponds to a pronounced multifractality on the level of larger fluctuations and its suppression
towards monofractality when going to small fluctuations. These small ones thus resemble a noisy background accompanying
the entire signal. Globally, such structures can easily be modeled as a superposition of independent, multifractally symmetric cascades.
Even in relation to reality such a modeling seems appropriate since participation of several somewhat independent multiplicative factors on a given phenomenon seems indeed possible. Out of those studied here the stock market index is an explicit sum
of prices of the individual companies entering the basket. The span of $f(\alpha)$ of such sums appears also visibly
narrower. This thus calls for caution when interpreting the width of $f(\alpha)$ as a measure of the 'degree of complexity'.
In case of asymmetric $f(\alpha)$ application of an appropriate multifractal cross-correlation
analysis~\cite{podobnik08,zhou08,oswiecimka14} in order to disentangle the original signal into uniform cascade components
may in this connection be recommended.
In the right-sided asymmetry the situation is reversed. Here, these are the smaller fluctuations
that develop a pronounced multifractal hierarchy while the largest ones are suppressed.
In relation to natural phenomena this seems to be a much rarer category of distortion.
Especially intriguing is the Sunspot Number variability case belonging to this category. The arguments presented here
indicate that possibly the famous Wolf formula~\cite{wolf1859} commonly used to digitally express this aspect of the
Sun activity does not represent a faithful mapping for the large events and needs to be refined such that the character of
cascading from the smaller events be extended to the large ones. Of course, at the present stage another possibility
that the large events are governed by a somewhat different dynamics cannot be excluded either. In any case, resolving
this issue emerges as a great scientific challenge.

We thank Drs Frederic Clette and Laure Lefevre of the Royal Observatory of Belgium, Brussels,
for very helpful exchanges on the issue of the Sunspot Number variability.

\end{document}